\documentclass{llncs}

\usepackage[utf8]{inputenc}
\usepackage[linesnumbered,vlined]{algorithm2e}
\usepackage{graphicx}
\usepackage{tabularx}
\usepackage{multicol}
\usepackage{multirow}
\usepackage{float}

\begin{document}

\title{RAdNet-VE: An Interest-Centric Mobile Ad Hoc Network for Vehicular Environments}
\titlerunning{RAdNet-VE}  % abbreviated title (for running head)
\author{Fabrício B. Gonçalves\inst{1, 2, 3} \and Felipe M. G. França\inst{1} \and Cláudio L. Amorim\inst{2}}
\authorrunning{Ivar Ekeland et al.} % abbreviated author list (for running head)
\tocauthor{Ivar Ekeland, Roger Temam, Jeffrey Dean, David Grove,
Craig Chambers, Kim B. Bruce, and Elisa Bertino}
\institute{Laboratório de Arquitetura de Computadores e Microeletrônica \\PESC/COPPE/UFRJ -- Rio de Janeiro, Brazil \and
Laboratório de Computação Paralela e Sistemas Móveis \\PESC/COPPE/UFRJ -- Rio de Janeiro, Brazil \and
Núcleo de Sistemas Complexos -- IFF-BJI/IFF -- Bom Jesus do Itabapoana, Brazil \\ 
e-mail: fabricio.goncalves@iff.edu.br, \{felipe, amorim\}@cos.ufrj.br
}

\maketitle              % typeset the title of the contribution

\begin{abstract}

In this study, we propose a variation of the RAdNet for vehicular environments (RAdNet-VE). The proposed scheme extends the message header, mechanism for registering interest, and message forwarding mechanism of RAdNet. To obtain results, we performed simulation experiments involving two use scenarios and communication protocols developed from the Veins framework. Based on results obtained from these experiments, we compare the performance of RAdNet-VE against that of RAdNet, a basic content-centric network (CCN) using reactive data routing, (CCN$_R$), and a basic CCN using proactive data routing, CCN$_P$. These CCNs provide non-cacheable data services. Moreover, the communication radio standards adopted in the scenarios 1 and 2 were respectively IEEE 802.11n and IEEE 802.11p. The results shown that the performance of the RAdNet-VE was superior to than those of RAdNet, CCN$_R$ and CCN$_P$. In this sense, RAdNet-VE protocol (RVEP) presented low communication latencies among nodes of just 20.4ms (scenario 1) and 2.87 ms (scenario 2). Our protocol also presented high data delivery rates, i.e, 83.05\% (scenario 1) and 88.05\% (scenario 2). Based on these and other results presented in this study, we argue that RAdNet-VE is a feasible alternative to CCNs as information-centric network (ICN) model for VANET, because the RVEP satisfies all of the necessary communication requirements.

\end{abstract}

\section{Introduction}

In recent years, vehicular ad hoc networks (VANETs) have received considerable attention from both industry and academia regarding potential applications for gathering, processing, and distributing security information, traffic conditions, and entertainment data. VANETs are a type of mobile ad hoc network (MANET) in which vehicles communicate with other vehicles and network facilities on highways, roads, or streets using wireless communication devices. However, VANETs have intermittent connectivity, highly dynamic topologies, and constant changes of density, making communication between nodes difficult.

Many methods have been proposed to facilitate communication between nodes in VANETs \cite{Lee:2010}\cite{Sharef:2014}, but their use of protocol stack models and node addressing schemes designed for Internet Protocol (IP)-centric networks makes them inadequate for highly dynamic vehicular environments \cite{Yu:2013}. Since source nodes need to know the addresses of the destination nodes to establish end-to-end communication and discover routes, they incur a message overhead to find and update their routing tables. Communication between nodes tends to be intermittent because of their high mobility. Therefore, Bai and Krishnamachari  \cite{Bai:2010} have argued for a paradigm shift in vehicular networking to develop information-rich applications.

To this end, some researchers have identified information-centric networks (ICNs) \cite{Ahlgren:2012} as a key paradigm, because they offer an attractive solution for highly mobile and dynamic environments such as VANETs. Among the architectural models found in the ICN literature, the content-centric network (CCN) has gained prominence in work on vehicular networking  \cite{Arnould:2011}\cite{Amadeo:2012a}\cite{Amadeo:2012b}\cite{Amadeo:2013}\cite{Wang:2012a}\cite{Wang:2012b}. Although CCN is more promising for vehicular environments than IP-centric models, there are some limitations preventing its adoption in VANET projects. For example, the flooding of interest packets due to interest packet-forwarding policies, whereby packets are forwarded to all of a node's neighbors at every new hop, can cause a broadcast storm. Moreover, the CCN model uses data structures such as routing tables, and adopts algorithms such as the ad hoc on-demand distance vector (AODV) \cite{Perkins:1994}, dynamic source routing (DSR) \cite{Jhonson:1996}, and greedy perimeter stateless routing (GPSR) \cite{Karp:2000}. These algorithms are vulnerable in highly dynamic vehicular environments because of the inherent path intermittence \cite{Yu:2013}. Finally, although there have been promising studies in the field of vehicular networking, these have only focused on scenarios regarding popular sharable data services \cite{Yu:2013}. Consequently, scenarios in which applications need to exchange a large amount of delay-sensitive data have not been studied. Applications with this feature make use of non-cacheable data services \cite{Yu:2013}. 

In this study, we propose a new ICN model for VANETs. This model is a variant of RAdNet (Inte\textbf{R}est-Centric Mobile \textbf{Ad} Hoc \textbf{Net}work) \cite{Dutra:2012} for \textbf{V}ehicular \textbf{E}nvironments (RAdNet-VE). In our model, each node uses an Active Prefix to compensate for the absence of an IP-centric mechanism, eliminating the need to uniquely identify nodes and the maintenance overhead due to routing information in the network. This Active Prefix is a simple data structure implemented in the network layer of each node, and is composed of a node prefix and an application interest. The node prefix is used for node identification, message addressing, and probabilistic message forwarding, whereas the application interest is used for name searching and group formation in a distributed manner. Interests are terms that have some meaning for applications, such as data and traffic events. Moreover, our model extends certain RAdNet data structures and mechanisms: the message header, the mechanism to register interest in the network layer, and the message forwarding mechanism. Thus, the communication protocol of RAdNet-VE (or RAdNet-VE protocol, RVEP) can forward messages according to the source prefix, interest, source relative position, and propagation direction.Further, RVEP can limit the scope of communication according to the number of hops registered with interest in the network layer of the nodes and the identifier of the road in which a node is placed, and can implement membership services in a distributed manner using interests in network messages. The nodes of RAdNet-VE do not cache data and do not perform in-network processing. We designed RVEP based on the communication requirements of application categories for VANETs proposed by Willke et al. \cite{Willke:2009}. 

We performed simulation experiments involving usage scenarios and communication protocols from the Veins framework \cite{Veins:2013}. We designed two scenarios to compare the performance of RAdNet-VE against that of RAdNet \cite{Dutra:2012}, a basic CCN using reactive data routing (CCN$_R$) \cite{Amadeo:2013} and a basic CCN using proactive data routing (CCN$_P$)\cite{Wang:2012b}. Both the CCN$_R$ and CCN$_P$ provide non-cacheable data services. In the first scenario, we simulated the cooperation among vehicles and semaphore signals using a three-by-three grid, where the distance between any two of the 16 intersections was 300 m. In this grid, each of eight corridors received a flow of 1500 veh/h. Cooperation allows the semaphore signals to collect data about incoming and outgoing vehicles at intersections, and adjust the phase times according to the flow of vehicles in upstream segments of the intersections. In this scenario, we adopted a radio model based on IEEE 802.11n, because the IEEE 802.11 standard has been used to deploy access points at intersections to provide Internet access and capture traffic data that may aid the routing of messages along a VANET \cite{Li:2007}. In the second scenario, we simulated the cooperation among vehicles once they had received information regarding an obstacle on the road. In this scenario, we built a 10 km road segment and placed a node on the roadside to notify vehicles about the presence of an obstacle at the end of the road segment. After receiving this notification, the vehicles must initiate cooperation among themselves to avoid collisions due to abrupt changes in speed in the vicinity of the obstacle. During the simulations for this scenario, the road segment received a flow of 1500 veh/h. Each vehicle was traveling at 80 km/h until reaching the obstacle at the end of the road segment. When crossing the obstacle region, each vehicle reduced its speed to 10 km/h. In this second scenario, we adopted a radio model based on IEEE 802.11p, because nodes equipped with IEEE 802.11p radios can communicate over a maximum range of 1000 m \cite{Ploeg:2013}. 

In summary, the main contributions of our research are as follows:

\begin{itemize}
	\item The proposal of a new ICN in which the data exchange is based only on the interests of applications for VANETs;
    \item We extended both the data structures and mechanisms of the RAdNet to create a communication protocol to satisfy the communication requirements of application categories for VANETs;
    \item Demonstrations of the feasibility of our RAdNet-VE as a VANET whose communication is centered on the interests of applications. These demonstrations resulted from an analysis of the simulation results from two distinct scenarios. In the first scenario, we evaluated the cooperation among vehicles and semaphore signals using a radio model based on IEEE 802.11n. In the second scenario, we evaluated the cooperation among vehicles using a radio model based on IEEE 802.11p.

\end{itemize}

This report is organized as follows. In Section 2, we present the works related to this study. In Section 3, outline the background to the research presented in this study. In Section 4, we describe the design of RAdNet-VE and its communication protocol. In Section 5, we describe two use scenarios and simulation settings. In Section 6, we present an analysis of the results obtained from the simulation experiments. In Section 7, we present a discussion regarding the results obtained from the simulation experiments. Finally, in section 8, we present our conclusions regarding this study.  

\section{Related Works}

There have been several studies on CCNs for VANETs. Most prominent are those conducted by Arnould \textit{ et al}. \cite{Arnould:2011}, Amadeo \textit{et al}. \cite{Amadeo:2012a} \cite{Amadeo:2013} and Wang \textit{et al} \cite{Wang:2012a}\cite{Wang:2012b}. 

Arnould \textit{et al.} \cite{Arnould:2011}, applied a CCN model to disseminate critical information in a hybrid VANET. Their study proposed an active data delivery mechanism, named the event packet, which does not require the prior sending of an interest packet in order that a data delivery occurs. The publisher detects critical events using sensors embedded in a vehicle and broadcasts event packets containing information related to delay-sensitive events such as accident information, safety alerts, and collision warnings. However, the authors modified the original CCN architecture to control the dissemination of event packets according to the bandwidth they require. This increased complexity may cause operational failures in high-demand environments.

Amadeo et al. \cite{Amadeo:2012a} \cite{Amadeo:2013} proposed the CRoWN architecture \cite{Amadeo:2012a} and content-centric vehicular networking (CCVN) \cite{Amadeo:2013}. They proposed a CCN-based framework for VANETs, and evaluated its performance using the IEEE 802.11p standard. According to Amadeo et al.  \cite{Amadeo:2012a}, CCN-based VANETs exhibit better performance than IP-based VANETs in terms of data transmission and the load balance of vehicles in the network, and suffer less performance degradation as the data volume increases. Moreover, the authors divided the interest packets into two sub-types basic interests (B-Int) and advanced interests (A-Int). B-Int was sent when a consumer wished to discover content and requested the first segment, whereas A-Int was used to request subsequent content from previously discovered providers. Moreover, the authors introduced a new data structure named the content provider table (CPT) to replace the forwarding information base (FIB). The CPT stores information regarding providers that have already been discovered and associates the MAC address of these nodes with the content. Thus, the authors discarded the main CCN premise, which is the independence of content from its physical location. This conceptual rupture of the original CCN proposal may compromise the support of the mobility of the nodes.

Wang \textit{et al.} \cite{Wang:2012a} presented a CCN architecture, but did not consider which applications or events would significantly affect its design. Moreover, the proposed architecture cannot be used efficiently in applications based on vehicle-to-vehicle communications. Wang \textit{et al.} \cite{Wang:2012b} also proposed a packet dissemination mechanism to reduce the latency of content delivery. This mechanism employed timers to coordinate the packet sending among the network nodes. However, the proposal of the mechanism did not establish the limits of interest dissemination within a geographical area. Thus, the flooding problem persisted. Finally, their evaluation did not consider how the mobility of the nodes would impact the proposed mechanism.

The main gap in the CCN literature is the absence of studies that consider scenarios based on applications that use non-cacheable data services \cite{Yu:2013}. Therefore, the studies described in this section have provided the background to build two types of CCN with non-cacheable data service: (i) CCN$_R$: this is a basic implementation of a CCN with non-cacheable data service using reactive data routing \cite{Amadeo:2013}. (ii) CCN$_P$: this is a basic implementation of a CCN with non-cacheable data service using proactive data routing \cite{Wang:2012b}. We created these CCNs to compare their performance with that of our proposed RVEP.

\section{Background}

In this section, we outline the background to the research presented in this study. We first describe the theoretical framework on which we designed RAdNet-VE. We present both the architectural model and the data structures and communication protocol of RAdNet \cite{Dutra:2012}. Finally, we describe four categories of communication requirements for VANET applications \cite{Willke:2009}. These categories present specific constraints on the message delivery latency and reliability, scale, scope of communication, and structure of communication groups  \cite{Willke:2009}.

\subsection{Interest-Centric Mobile Ad Hoc Network}

RAdNet is based on the publisher/subscriber architectural model \cite{Buschmann:1996} in which a publisher sends a message with some domain of interest to all nodes in the network, and the subscriber nodes of this specific interest receive the message. This communication model differs from traditional approaches, where information in the network has a source and a destination. By using the interest mechanism in messages, RAdNet implements the publisher/subscriber model in a completely distributed manner.

The publisher/subscriber model was initially developed in RAdNet for MANETs, because incoming and outgoing devices are frequent in such networks. When a device moves, it can enter or leave the transmission range of its neighbors. Such network instability does not impact RAdNet, because the nodes do not need to know the network topology to send messages, as is the case in most routing protocols for MANETs. Moreover, nodes do not require a unique address, because communication is based on interest.

Interest refers to any term (or sequence of characters) that has a meaning for applications, allowing the information to focus on the application rather than the device. The use of interest reduces the message overhead and obviates the need for routing tables, because messages with interest do not have a unique and predetermined destination. Message forwarding occurs from a set of probabilistically chosen fields, as in a gossip protocol, hence reducing the complexity of forwarding decisions. In RAdNet, nodes need not determine which neighbor receives a message. Prior to forwarding messages, each node makes its own forwarding decision using criteria defined by a matching function.

\begin{figure}[t]
	\centering
	\includegraphics[scale=0.75]{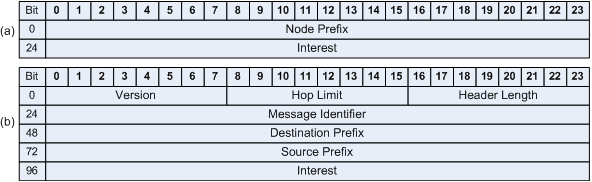}
	\caption{RAdNet data structures: (a) Active Prefix and (b) Message Header}
	\label{fig:radnet_headers}
\end{figure}

In RAdNet, each node defines its Active Prefix (see Fig.  \ref{fig:radnet_headers} (a)), which is divided into two parts. The first is called the Node Prefix, and is composed of probabilistically chosen fields. The second is called the Interest, and contains the user or application interest. The fields of the Node Prefix are used for two functions: (i) probabilistic message forwarding, where each node has a forwarding probability given by Prefix building; and (ii) addressing, where a bit sequence of the Prefix fields identifies the node. The message headers contain a RAdNet protocol version, hop limit, header length, message ID, two node prefixes that identify the associated source and destination nodes, and the application interest (see Fig. \ref{fig:radnet_headers} (b)).

The RAdNet node generates its prefix with $n$ fields of $m$ bits, in such a way that the $n \times m$ bits provide node identification for addressing purposes as well as a matching filter for message forwarding. Specifically, a node generates a sequence of $n$ fields where, for each field, the node assigns a value with $m$ bits using a random variable with some probability distribution. The resulting set of field values constitutes the node's prefix, and the node is identified by concatenating the field values in the same order.

Each RAdNet node runs a communication protocol that allows nodes to forward messages according to the matching result among their active prefixes and the received specific header fields of the RAdNet messages. These specific fields are the Source Prefix and Application Interest. Algorithm \ref{alg:radnetprotocol} describes the protocol processes involved in forwarding a RAdNet message.

\begin{algorithm}[t]
  \SetKwData{Msg}{Msg}
  \SetAlgoLined
  	\scriptsize \KwIn{msg$_j$}
  	
  	\eIf{$msg_j.ID \in idTable_{i}[msg_{j}.srcPrfx]$} {
	  \textbf{Discard} msg$_j$\;
  	}{
	  \textbf{Insert} msg$_j$.ID \textbf{into} idTable$_{i}$[msg$_{j}$.srcPrfx]\;
	  
      msg$_j$.hopLimit := msg$_j$.hopsLimit - 1\;
      intrstMtch := msg$_{j}$.appIntrst $\in$ intrstTable$_{i}$\;
	  prfxMtch := $|$prefix$_i \cap$ msg$_j$.srcPrfx$| > 0$  $\vee$ msg$_j$.destPrfx =  \textbf{null}\;
	  
	  \If{intrstMtch = \textbf{true}}{
	    \If{$msg_{j}.destPrfx$ = \textbf{null} $\vee msg_{j}.destPrfx = prefix_i$}{
	      \textbf{Send } a copy of $msg_j$ \textbf{to} application\;
	    }
	  }
	  
	  \eIf{prfxMtch = \textbf{true} $\vee$ intrstMtch = \textbf{true}}{    
	      \eIf{msg$_j$.hopLimit $>$ 0} {
		  \textbf{Send } msg$_j$ \textbf{to} \textbf{all} $neighbors_i$\;
	      }{
			\textbf{Discard} msg$_j$\;
	      }     
	  }{
	    \textbf{Discard} msg$_j$\;
	  }
	  
  	}
  \caption{RAdNet communication protocol}
  \label{alg:radnetprotocol}
\end{algorithm}

Upon receiving a RAdNet message ($msg_j$) from any neighbor $j$, the algorithm checks whether node $i$ has already received such a message. If so, node $i$ discards $msg_j$; otherwise, it inserts the ID of the RAdNet message and its source prefix into the table of identifiers. Moreover, the algorithm reduces the hop limit of the RAdNet message. The algorithm then performs message filtering. It first checks whether the table of interests of node $i$ has an entry equal to the interest of the RAdNet message ($msg_{j}.appIntrst$) and then checks whether the prefix of node  $i$'s prefix ($prefix_i$) and the source prefix of the RAdNet message ($msg_j.srcPrfx$) have one or more matching pairs of fields, or if the destination prefix is $null$. If the interests match, the algorithm checks whether node $i$ is the destination of $msg_j$. If so, the algorithm creates a copy of $msg_j$ and forwards it to the application associated with $msg_{j}.appIntrst$. Finally, if the prefix or interest match and the hop limit of the message  ($msg_j.hopLimit$) is greater than zero, node $i$ forwards $msg_j$ to its neighbors ($neighbors_i$). Otherwise, $msg_j$ is discarded.

\subsection{Communication Requirements of Application Categories for VANETs}

Based on their communication requirements, VANET applications can be classified into four categories: general information services, security information services, individual movement control, and group movement control. These categories represent specific constraints on message delivery latency and reliability, scale, the scope of communication, and the structure of communication groups \cite{Willke:2009}. Message delivery latency and reliability are critical performance measures. Scale is important because scenarios can involve many vehicles. The scope of communication drastically affects the scalability of the application, because it depends on the manner in which messages are forwarded and the network is organized. The structure of the communication group refers to the capability of vehicles to establish persistent relationships or communicate with other vehicles.

Applications belonging to the category of general information services can tolerate delays in message delivery. These applications can also tolerate intermittent communication failures. With regard to scale, such applications often transmit messages to a large area, and hence require a large scope of communication. As general information applications do not address vehicle movement control, they do not need to maintain structures for group communication among vehicles.

A key requirement of security information applications is hard real-time operation. As a result, such applications can fail when there are delays in message delivery. These applications require short processing times and latencies of 40 ms, as well as a message sending frequency of 50 Hz. Furthermore, they need high message delivery rates. The scale, scope of communication, and structure requirements of the communication group pertaining to security information applications are the same as those of general information services.
As for security information applications, those related to individual movement control have hard real-time constraints, but can tolerate failures due to infrequent delays in message delivery. Moreover, these applications use data in the neighborhood of vehicles to ensure driver security and maintain the optimal distance among vehicles. Individual movement control applications are middle-scale VANETs, and so their scope of communication is smaller than that of general information and security information applications. With regard to the structure of the communication group, individual movement control applications do not involve persistent groups, but form transient relationships among vehicles.

In applications related to group movement control, message delivery latency can vary according to the movement regulation model. For example, in models for group planning with separate regulations, applications can tolerate message delivery delays with no failures, because the real-time constraints are more stringent than in other models. With regard to the reliability of message delivery, group movement control applications should be able to determine whether messages have been received by the appropriate vehicles. The vehicles can then act appropriately if messages are not delivered within the stipulated time. The scale used in group movement control applications is similar to that for individual movement control. Thus, the scope of communication is limited to the neighborhood of a vehicle or similarly small regions. Unlike other VANET applications, group movement control has a persistent communication structure, because vehicles maintain relationships with other specific vehicles.

\section{Interest-Centric Vehicular Ad Hoc Network}

Understanding the communication requirements of application categories for VANETs is critical in designing an efficient communication protocol. In this sense, we initially show how our RAdNet-VE design addresses such requirements. Next, we describe how the design of RAdNet-VE extended the message header of the RAdNet. We also describe the design of the communication protocol of the RAdNet-VE. Finally, we describe how RAdNet-VE is related to the short-range communication radio technologies.

\subsection{Addressing the Communication Requirements of Application Categories for VANETs}\label{sec:addressing}

Communication protocols for VANETs should take into account the needs of latency, reliability, and scale. The scope and services of group members should also be well-defined to satisfy the communication requirements of the VANET application categories \cite{Willke:2009}. The RAdNet-VE inherits the characteristics of RAdNet, and adopts the following mechanisms and approaches.

When an application needs to communicate at low latency, the communication protocol should be able to provide end-to-end communication with little delay \cite{Willke:2009}. In this sense, RAdNet-VE should provide low latency, because it does not suffer under the dynamism of VANETs. In RAdNet-VE, nodes do not require information on the network topology. Therefore, they do not need to find, keep, and update routes for other nodes in the network. As a consequence, the bandwidth normally occupied by control messages is released. In RAdNet, the use of prefixes as a set of probabilistic values allows messages to be forwarded according to the probability distribution used to build the prefixes. Thus, the nodes do not need to determine the best path between the source and the destination of a message. The mechanism adopted by RAdNet allows for multiple paths to exchange messages among vehicles.
  
Applications require a protocol that can deliver messages to a group of nodes \cite{Willke:2009}. This protocol should ensure a high probability of message delivery \cite{Willke:2009}. As in RAdNet nodes, RAdNet-VE nodes store the message ID and source prefix for comparison with those of the received messages. This allows duplicate messages, which should not be forwarded, to be detected. However, this is not enough to ensure high message delivery rates in RAdNet-VE, because vehicular environments feature high vehicle densities. Thus, when vehicles in the same neighborhood receive a message, they forward it to all neighbors according to a RAdNet matching filter, causing an unnecessary message overhead. To avoid this, we extend the message forwarding mechanism of the RAdNet by adding a field to the original RAdNet message header. This field should store the relative position of message source, direction and road identifier. In this study, a road identifier can identify a highway, road or street and defines the geographic area where nodes (i.e., vehicles and road-side units) are operating. Moreover, we also took into account the use of the global positioning system (GPS) devices within each RAdNet-VE node and the access to map databases from applications for VANETs. Therefore, when a node receives a message, it can calculate the relative distance between it and the source of the received message, and store both the prefix and relative distance of the source of the received message. It is important to point out that the nodes should only store the prefixes and the relative distances of neighbor nodes. As a result, each node should know the relative positions of all of its neighbors. Therefore, if a node is farther from a message source, it should forward the message to its neighbors. Otherwise, it will act passively and not forward the message. The addition of these constraints to the message forwarding mechanism should allow messages to be delivered to many nodes using few hops, and should provide low latency communication among nodes.

The use of relative positions to forward messages over long distances should allow RAdNet-VE to scale appropriately in high-density vehicular environments \cite{Willke:2009}. However, some applications for VANETs need to propagate messages in a given direction. Although RAdNet nodes can forward messages over long distances using many hops, the message forwarding mechanism of the RAdNet does not satisfy such requirement, because it was designed to satisfy the requirements of applications for MANETs. Therefore, we modified the message forwarding mechanism of the RAdNet to address this issue by adding a direction field to the message header of the RAdNet. This field admits the following three values:
 
\begin{itemize}
 \item -1: messages can only be forwarded in the direction opposite to that of the node;   
 \item 0: messages can be propagated in all directions; 
 \item 1: messages can only be forwarded in the same direction as the node is moving.
\end{itemize}

The addition of the direction field in the header of the RAdNet message allows message forwarding flow to be unidirectional or bidirectional. Hence, we add one more constraint to the message forwarding mechanism, which should allow each node to forward messages according to the direction in which it moves. When receiving a message, the node should use the relative position field to calculate its position in relation to the message source. Since each node owns its own GPS device, it should be able to know its current positioning in relation to their neighbors. In other words, nodes should be able to know whether they are behind or in front of a message source. To represent the positioning of nodes, we adopt the value of -1 when the node is behind the message source and 1 when it is in front of it. Thus, when receiving a message whose source is in front of it, if the direction field is -1, the node should forward this message to its neighbors; otherwise, it should discard the message. On the contrary, when receiving a message from a source behind of it, if the direction field is 1, the node should forward this message to its neighbors; else, it should discard the message. Finally, the nodes should also forward messages whether the road identifier corresponds to the way where nodes  are operating.

Applications related to group and individual movement control operate in a well-defined communication scope, which can be a neighborhood of vehicles, or a  small region in the network \cite{Willke:2009}. Thus, communication protocols must ensure selective message delivery, which can be based on trajectory, vehicle proximity, or vehicle identification \cite{Willke:2009}. In RAdNet, nodes do not need to maintain and update routes to other nodes, but they can forward messages until a maximum number of hops according to the result of the matching filter \cite{Dutra:2012}. However, such behavior does not allow applications to control individual or group movements of vehicles, because RAdNet takes into account a unique value of maximum number of hops to forward messages. Thus, we address this question by extending the method of registering interests in the node and the message forwarding mechanism. In RAdNet-VE, applications for VANETs should register interests with the respective maximum numbers of hops. We extended the message forwarding mechanism by using the road identifier field. In this sense,  the communication scope is restricted to the maximum numbers of hops registered with each interest and to the ways where nodes are operating.

Regarding services to group members, applications related to the control of group or individual movements need a protocol that enables persistent group structures to be maintained and updated. Since RAdNet-VE is an ICN, it does not consider the group addressing space or distributed/centralized mechanisms for group member management. As a RAdNet, RAdNet-VE is based on the publisher/subscriber architectural model (i.e., a publisher sends a message with interest to all nodes in the network, and the subscriber nodes of this interest receive the message using any of a variety of traditional approaches where the information transmitted over the network has a source and a destination). Using the interest mechanism in a message, RAdNet-VE should implement services for group members in a completely distributed manner.

\subsection{Description of the Extensions of the Message Header of the RAdNet}

Our proposed protocol extends certain RAdNet data structures and mechanisms. These extensions can be described as follows:

\begin{figure}[t]
	\centering
	\includegraphics[scale=0.75]{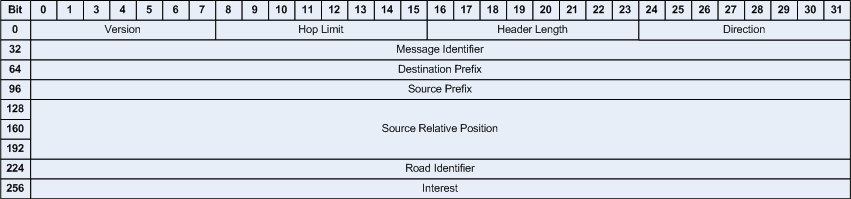}
	\caption{RAdNet-VE message header}
	\label{fig:radnet_ve_header}
\end{figure}

The extension proceeded as follows:
\begin{itemize}
    \item An increase from 24 to 32 bits in the length of the following message header fields: message identifier, destination prefix, source prefix, and interest. We increased the length of these fields to standardize them to base-2 values. These changes are illustrated in Fig. \ref{fig:radnet_ve_header}.
    \item The addition of three fields to the message header: the relative position of the message source (96 bits), the direction of message forwarding (8 bits), and the road identifier (32 bits). These new fields and their respective lengths are illustrated in Fig. \ref{fig:radnet_ve_header}. In this study, we assume that all nodes are equipped with GPS devices, and that applications running on these nodes can access map databases such as Open Street Map \cite{OSM:2016} and Google Maps \cite{Google:2016}. Moreover, the relative position of the message source must record 32-bit longitude, latitude, and altitude values. 
    \item The addition of new rules extending the message forwarding mechanism. It is important to point out that the addition of new rules for message filtering does not affect those of the RAdNet communication protocol.  
\end{itemize}

These extensions enable RVEP to forward messages by taking into account the following data: (i) the identifier of the way in which nodes (vehicles and RSUs) are operating, (ii) the result of the matching between the node prefix and source prefix of a message, (iii) the result of the matching between the interests registered in the nodes and those contained in the message, (iv) the distance between a node and the message source, (v) the position of the node in relation to the message source, and (vi) the maximum numbers of hops registered with interests in the network layer of the nodes or a default maximum number of hops, which is only used whether the interest in the messages do not match with the interests registered in the network layer of the node.

\subsection{Design of the RAdNet-VE Communication Protocol}

In RAdNet-VE, all nodes must be initialized before they can transmit messages. During the node initialization stage, the following control variables and data structures are initialized: sent message counter, node prefix, table of source prefixes and received message identifiers, table of interests and maximum number of hops, table of relative positions of neighbors within range of radio communication, and road identifier list. Following the initialization process, the node registers the interests of applications with their maximum number of hops. Finally, if the node is a road-side unit, the list of road identifiers may contain more than one entry, because it may be positioned alongside more than one road. Thus, if the node is responsible for notifying vehicles about obstacles on the road, the list of road identifiers should have one entry, so that the node publishes data regarding the conditions of a unique roadway. However, if the node is responsible for controlling the traffic lights at a complex intersection, the list of road identifiers should have entries for each street upstream of the intersection. If the node is a vehicle, the list of road identifiers should have a single entry that is updated when the node moves onto a new path.

\begin{algorithm}[!t]
  \SetKwData{Msg}{Msg}
  \SetAlgoLined
  	\scriptsize \KwIn{msg$_j$}
  	\eIf{$msg_j.ID \in idTable_{i}[msg_{j}.srcPrfx]$} {
	  \textbf{Discard} msg$_j$\;
  	}{
	  \textbf{Insert} msg$_j$.ID \textbf{into} idTable$_{i}$[msg$_{j}$.srcPrfx]\;
	  msg$_j$.hopLimit := msg$_j$.hopsLimit + 1\;
      \If{msg$_j$.hopLimit = 1}{
      	\textbf{Insert} msg$_j$.position \textbf{into} posTable$_{i}$[msg$_{j}$.srcPrfx]\;
      }
	  \eIf{msg$_j$.roadId $\in$ roadIdentifiers$_i$}{
      	positioning$_i$ := calcPositioning(position$_i$, msg$_j$.position, msg$_j$.roadId)\;
        fwdPrfx := prefix$_i$\;
        dist := calculateDistance(position$_i$, msg$_j$.position, msg$_j$.roadId)\;
        \ForEach{position $\in$ posTable$_i$ } {
          \If{calcPositioning(position, msg$_j$.position, msg$_j$.roadId) = msg$_j$.direction}{
              neighborDist := calculateDistance(position, msg$_j$.position, msg$_j$.roadId)\;
              \If{neighborDist $>$ dist $\wedge$ neighborDist $\leq$ radioRange/2}{
                  fwdPrfx := Neighbor's prefix\;
                  dist := neighborDist\;
              }
          }
        }
        intrstMtch := msg$_{j}$.interest $\in$ intrstTable$_{i}$\;
        prfxMtch := $|$prefix$_i \cap$ msg$_j$.srcPrfx$| > 0$  $\vee$ msg$_j$.destPrfx =  \textbf{null}\;
        \If{intrstMtch = \textbf{true} $\wedge$ (positioning$_i$ = msg$_j$.direction $\vee$ msg$_j$.direction = 0) }{
          \If{msg$_{j}$.destPrfx = \textbf{null} $\vee$ msg$_{j}$.destPrfx = prefix$_i$}{
            \textbf{Send } a copy of msg$_j$ \textbf{to} application\;
          }
        }
        \eIf{prfxMtch = \textbf{true} $\vee$ intrstMtch = \textbf{true}}{
          fwdMsg := msg$_{j}$.destPrfx = \textbf{null} $\vee$ msg$_{j}$.destPrfx $\neq$ prefix$_i$\;
          nodePos := fwdPrfx = prefix$_i$ $\wedge$ (positioning$_i$ = msg$_j$.direction $\vee$ msg$_j$.direction = 0)\;
          fwdHops := \textbf{false}\;
          \eIf{intrstMtch = \textbf{true}}{
              fwdHops := msg$_j$.hopLimit $<$ interstTable$_i$[msg$_{j}$.interest]\;
          }{
              fwdHops := msg$_j$.hopLimit $<$ maxHopLimitOfProtocol\;
          }
          \eIf{(fwdMsg $\wedge$ nodePos $\wedge$ fwdHops) = \textbf{true}}{
              \textbf{Wait } uniform(0, 1)/dist\;
              \textbf{Send } msg$_j$ \textbf{to} \textbf{all} neighbors$_i$ \;
          }{
              \textbf{Discard } msg$_j$\;
          }
        }{
          \textbf{Discard } msg$_j$\;
        }
      }{
      	\textbf{Discard} msg$_j$\;
      }
  	}
  \caption{RAdNet-VE communication protocol}
  \label{alg:radnetveprotocol}
\end{algorithm}

For the node to send a message, the communication protocol must receive three items of data: destination prefix, application interest, and message direction. Before sending the message to its neighbors, the node builds the message as follows: (i) configure the version field with the version of the protocol at the given time; (ii) set the number of hops to 0, so that the value of this field increases as the message is forwarded by other nodes; (iii) configure the length of the message header with an equivalent integer value; (iv) configure the identifier field with the value of the message counter at the given time; (v) configure the destination prefix field with the corresponding input data; (vi) configure the source prefix field with the prefix node; (vii) configure the application interest field with the corresponding input data; (viii) configure the position field with data obtained from the GPS device embedded in the vehicle; (ix) configure the direction field with the corresponding input data; and (x) configure the road identifier field with the corresponding input data. After sending the message, the sent message counter increases by one.

Upon receiving $msg_j$, node $i$ executes the Algorithm \ref{alg:radnetveprotocol}.The algorithm first checks whether its identifier ($msg_j.id$) exists in the table of source prefixes and received message identifiers ($idTable_i$). If it so, the node discards $msg_j$; otherwise, the node registers $msg_j.id$ in $idTable_i$, and increases the hop limit value of msg$_j$. If the number of hops in $msg_j.hopLimit$  is 1, the algorithm inserts the relative position of the message source into the table of neighbor relative positions ($posTable_i$). This allows the node to update the relative positions of neighbors that are one hop away. Next, the algorithm checks whether the road identifier value of $msg_j$ exists in the list of road identifiers. If not, the node discards $msg_j$; otherwise, it calculates the position of the node in relation to the message source. The result of this calculation should be -1 or 1, making it possible to determine whether the message source is behind or in front of the node.  Based on the relative positions in $posTables_i$, the node can determine whether it or any of its neighbors is farthest from the message source. The algorithm extracts this information from $posTables_i$ and stores it in  $fwdPrfx$.  In the next two steps, the algorithm performs message filtering inherited from RAdNet. It first checks whether the table of interests of node $i$ ($interestTable_i$) has an entry equal to interest ($msg_{j}.interest$). The algorithm then determines whether the source prefix ($msg_j.srcPrfx$) of the RAdNet-VE message has one or more matching pairs of fields, or if the destination prefix is $null$. If an interest match occurs, the algorithm checks whether node $i$ is the destination of msg$_j$, or if the destination prefix is $null$. If so, the algorithm creates a copy of msg$_j$ and forwards it to the application associated with $msg_{j}.appIntrst$.

Before forwarding messages, the algorithm checks whether prefix matching or interest matching has occurred. If not, it discards $msg_j$; otherwise, node $i$ waits for a short time before sending $msg_j$ to its neighbors (see Line 45). However, three conditions must be satisfied:

\begin{enumerate}
    \item node $i$ is not the destination of RAdNet-VE message;
    \item node $i$ is the farthest node from the message source;
    \item $msg_j$ has not moved the maximum number of hops. 
\end{enumerate}

If these conditions are not satisfied, the algorithm discards $msg_j$. For the first condition, the algorithm checks whether node $i$ is the destination of $msg_j$. When this condition is not satisfied, an application associated with $msg_j.appIntrst$ has received a copy of $msg_j$. Hence, the message should be discarded, because it has already reached its destination. For the second condition, the algorithm checks whether node $i$ is the farthest from the source of the message, and its position allows it to forward messages to its neighbors. If this condition is not satisfied, there is a node farther from the message source. For the third condition, the algorithm checks whether $msg_i$ has reached the maximum number of hops. When interest matching occurs, the algorithm checks whether $msg_j.hopLimit$ is lower than the maximum number of hops of the interest. Otherwise, it checks whether $msg_j.hopLimit$ is less than the maximum number of hops of the protocol. If one of these conditions is false, $msg_j$ discarded.

Finally, the algorithm determines the wait time through the result of a random number generation function of the uniform distribution between 0 and 1 divided by the distance between the node $i$ and the message source. The wait time is necessary, because it avoid that many nodes forward messages in closer instants. As the message moves away of its source, the wait time decreases.

\subsection{RAdNet-VE and Access Technologies}

\begin{figure}[t]
	\centering
	\includegraphics[scale=0.75]{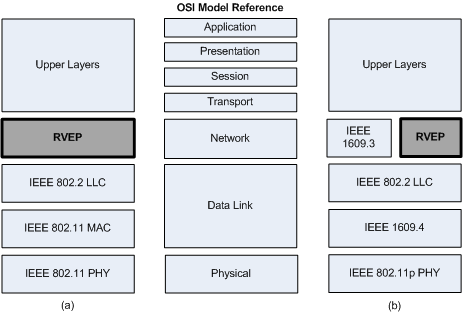}
	\caption{RVEP in the network layers: (a) RVEP as a replacement of IP in nodes equipped with radios based on IEEE 802.11, and (b) RVEP as a replacement of IP in nodes equipped with radios based on IEEE 802.11p}
	\label{fig:radnet_ve_radios}
\end{figure}

Once the RVEP has been designed, we need to specify how it runs on communication devices using short-range radio technologies. It is important to note that there are a variety of short-range radio technologies inter-vehicle communications, such as Bluetooth and Zigbee \cite{Sommer:2015}. However, research on vehicular networking has been dominated by approaches that were based first on communication radios based on the IEEE 802.11 standard (WiFi) \cite{IEEE80211:2012} and then on communication radios based on the IEEE 802.11p standard \cite{IEEE80211p:2010}. Although other radio communication technologies such as LTE (Long Term Evolution) and, white spaces and cognitive radio are used in vehicular networking,  \cite{Sommer:2015}, this study only takes into account short-range communication radio technologies. We will treat the other access technologies in future studies. 

We designed the RVEP to be a network layer protocol. In this sense, RVEP replaces IP in the two communication stacks shown in Figs. \ref{fig:radnet_ve_radios}(a) and Fig. \ref{fig:radnet_ve_radios}(b). 

As can be seen in Fig. 3(b), RVEP divides the network layer with component defined by IEEE 1609.3 standard. According to Sommer and Dressler \cite{Sommer:2015}, the IEEE 1609.3 standard was developed to support the provision and use of services on multiple channels and is part of the IEEE 1609 standard. The IEEE 1609 standard is also called Wireless Access in Vehicular Environments (WAVE) \cite{Sommer:2015}. Furthermore, IEEE 1609 Standard defines a complete Intelligent Transportation System (ITS) stack over IEEE 802.11p \cite{Sommer:2015}. However, this study does not address issues related to the dissemination of WAVE Service Advertisements (WSAs) and data exchanges through WAVE Short Messages (WSMs). 

\section{Experiments}

To test our proposal, we used the Veins framework \cite{Veins:2013}  implement RVEP in two scenarios. We also used the framework to implement the communication protocol of RAdNet (RP)  \cite{Dutra:2012}, CCN$_R$ and CCN$_P$. The Veins framework allows VANET simulations to be implemented on two simulators: (i) Omnet++ \cite{Omnet:2013}, which is an event-based network simulator; (ii) SUMO (Simulation of Urban Mobility) \cite{SUMO:2013}, which is a traffic simulator. We also used Veins to construct scenarios for testing each protocol . These scenarios are presented in the next subsection.

\subsection{Use Scenarios}

We constructed two scenarios to simulate the cooperation among nodes in ITS. In the first scenario, we considered a distributed traffic control system in which vehicles cooperate with semaphore signals. This cooperation allows the semaphore signals to collect data regarding incoming and outgoing vehicles at intersections, and adjust the phase times according to the occupancy rate of the upstream segments of the intersections. In this sense, we implemented two applications: (i) Traffic Light Controller (TLC): an application that runs on semaphore signals at intersections and varies the phase times according to the density of traffic in upstream segments; (ii) Driver Assistant (DA): an application that cooperates with TLC by sending data related to the arrival and departure of vehicles in upstream segments. In the second scenario, we considered an environment in which vehicles are searching for road condition information. When vehicles receive information related to an obstacle, they activate their cooperative adaptive cruise control to control their speed. For this purpose, we developed two further applications: (i) Cooperative Adaptive Cruise Control (CACC): an application that controls vehicles' speeds using information from neighboring vehicles; (ii) Obstacle Notifier (ON): an application that runs on road-side units and notifies vehicles of obstacles on the road.

In the following sections, we describe the cooperation of these applications in their respective scenarios.

\subsubsection{Scenario 1: Cooperation between vehicles and semaphore signals}\label{sec:usescenario1}

In this scenario, we created a three-by-three grid in which the distance between any two of the 16 intersections was 300 m. At each intersection, we installed a semaphore signal equipped with a wireless communication device to communicate and cooperate with vehicles that are approaching or leaving the intersection. Each signal was connected to those at the neighboring intersections.

Initially, each TLC requested data regarding the upstream segments of the intersection from all neighboring semaphore signals. Upon receiving this request, the TLCs sent a response to the requester. After receiving the response, the TLC extracted the data and calculated the offset of the other semaphore signals. Finally, each TLC initiated the collection of data regarding incoming and outgoing vehicles at the relevant intersection.

In the data collection stage, the TLCs attempted to determine the number of vehicles in upstream segments that are approaching an intersection with red lights. Hence, they requested data regarding vehicles approaching an intersection. After receiving this request, the DAs sent a response to the requester. The TLCs extracted the identifier of the DA instances from the responses, and stored them in an appropriate set of identifiers. From this data, the TLCs calculated the occupancy rate of upstream segments and used the highest occupancy rate to adjust the phase times of the semaphore signals. The TLCs also need to know the number of vehicles leaving the intersection. Therefore, they requested data regarding vehicles leaving an upstream segment at a green traffic light. After receiving this request, the DAs sent a response to the requester. From these responses, the TLCs extracted the DA instances identifier, and stored it in an appropriate set of identifiers. All TLCs requested data regarding incoming and outgoing vehicles in the intersection at 1 s intervals.

When a DA received a request from a TLC, it requested an acknowledgment of its previous response. After receiving this request, the TLC sent the acknowledgment to the DA requester. Once this response had been received, the DA did not respond to further requests for 5 s. We define this time to avoid DAs always response requests sent bt TLCs.

To adjust the phase times of the semaphore signals, the TLCs requested the occupancy rates calculated by their neighbors. After receiving this request, each TLC sent the highest occupancy rates and identifiers of the upstream segments of the relevant intersection to the requester. Upon receiving this response, the TLC extracted the occupancy rates and identifiers of the upstream segments and stored them in a dictionary. The TLCs repeated this process until the size of the dictionary was equal to the number of upstream segments in the three-by-three grid. Each TLC then calculated the phase times of the semaphore signal using the multiple edge reversals scheduling algorithm \cite{Barbosa:2000}. Moreover, each TLC adjusted the phase times of the semaphore signals every 300 s.

As the vehicles entered a new upstream segment, the DAs updated all data related to the particular segments in which their respective vehicles were traveling. This operation involved changing the identifier of the old upstream segment for that of the new upstream segment. According to the network protocol used by the nodes, the DAs used the upstream segment identifiers to reconfigure the network layer of the nodes. In this case, the DAs reconfigured the network layer of the nodes once the vehicles had entered the new road segment.

\subsubsection{Scenario 2: Cooperation between vehicles}\label{sec:usescenario2}

In this scenario, we created a 10 km road segment and installed an obstacle notifier into the road-side units of this segment. These units were equipped with a wireless communication device. The main goal was to notify vehicles of an obstacle at the end of the road segment. Moreover, we placed the road-side units such that vehicles were notified when they were 1 km from the obstacle. Finally, each vehicle traveled at 80 km/h until reaching the region of the obstacle. When crossing this region, each vehicle reduced its speed to 10 km/h.

To receive data related to obstacles on the road, the CACC sent a request to the ON. Upon receiving the request, the ON sent a response to the CACC requester, whereupon the CCAC initialized a neighborhood discovery process.

During the neighborhood discovery process, the CACC requested data related to all vehicles within communication range. When other CACCs received this request, they sent a response containing their identifier and position to the CACC requester. The CACC requester extracted the identifier and position of the neighbor, verified whether the neighbor was behind or in front, and stored the data in an appropriate data structure. Once the neighborhood discovery process had been initiated, it ran at 1 s intervals.

The CACC identified the following vehicle, and sent ten requests per second to the following CACC to obtain its speed and position. When a CACC received such requests, it responded by sending its speed and position to the CACC requester. The CACC requester extracted the speed and position of the following vehicle and calculated its own speed using the equations of Kato \textit{et. al} \cite{Kato:2002}.

\subsection{Simulation Settings}

We performed an extensive set of simulations to compare the performance of RAdNet-VE with that of RAdNet, CCN$_R$ and CCN$_P$. The simulation time was fixed to 3600 s, and all nodes were equipped with GPS devices and the applications described above. All applications were able to access a map database. 

\subsubsection{Network Parameters}

In scenario 1, we adopted the settings of IEEE 802.11n communication radios because the IEEE 802.11 Standard (WLAN) has been used to deploy access point in intersections to provide Internet access, capture traffic data or data that may aid in the routing of messages along a vehicular ad hoc network \cite{Li:2007}. The parameters of the IEEE 802.11n communication radios are listed in Table \ref{tab:radiosettings}. The MAC layer adopted the Carrier Sense Multiple Access with Collision Avoidance (CSMA/CA) protocol. The parameters for the CSMA/CA protocol are given in Table \ref{tab:macsettings}.

\begin{table}[h]
    \centering
    \caption{Settings of IEEE 802.11n communication radios used by network nodes in the simulations.}
    \label{tab:radiosettings}
    \scriptsize \begin{tabular}{|l|c|}
      \hline
      \textbf{Parameter} & \textbf{Value} \\
      \hline
      Transmission power & 158.48mW\\ \hline
      Signal attenuation threshold & -90dBm\\ \hline
      Carrier frequency of the channel & 5.0 GHz\\ \hline
      Thermal noise & -160dbm\\ \hline
      Sensitivity of the physical layer & -87dBm\\ \hline
      PHY header length & 128 bits \\ \hline
            Communication range & 200m \\ \hline
    \end{tabular}
\end{table}

\begin{table}[h]
    \centering
    \caption{Settings of IEEE 802.11n MAC used by network nodes in the simulations.}
    \label{tab:macsettings}
    \scriptsize \begin{tabular}{|l|c|}
      \hline
      \textbf{Parameter} & \textbf{Value} \\ \hline
      Length of the Queue & 100\\ \hline
      Slot duration & 0.0005s\\ \hline
      Difs & 0.00011s\\ \hline
      Number of transmissions attempts & 14\\ \hline
      Bitrate & 11.35 Mbps \\ \hline
      Contention window & 20\\ \hline
      MAC header length & 256 bits \\ \hline
    \end{tabular}
\end{table}

In scenario 2, we adopted the settings of IEEE 802.11p communication radios because nodes equipped with these radios can communicate into maximum range of 1000m. Furthermore, this radio model is most appropriate for scenarios where vehicles travel in high speed \cite{Ploeg:2013}. If there is an obstacle on the road, these vehicles need to be notified well in advance about its presence, so that they can initialize a cooperation. The parameters of the IEEE 802.11p communication radios are listed in Table \ref{tab:ieee80211psettings}. The MAC layer adopted CSMA/CA protocol. The parameters for CSMA/CA protocol are given in Table \ref{tab:ofdmsettings}.

\begin{table}[h]
    \centering
    \caption{Settings of IEEE 802.11p communication radios used by network nodes in the simulations.}
    \label{tab:ieee80211psettings}
    \scriptsize \begin{tabular}{|l|c|}
      \hline
      \textbf{Parameter} & \textbf{Value} \\
      \hline
      Transmission power & 200mW\\ \hline
      Signal attenuation threshold & -89dBm\\ \hline
      Carrier frequency of the channel & 5.89 GHz\\ \hline
      Thermal noise & -110dbm\\ \hline
      Sensitivity of the physical layer & -89dBm\\ \hline
      PHY header length & 46 bits \\ \hline
      Communication range & 1000m \\ \hline
    \end{tabular}
\end{table}

\begin{table}[h]
    \centering
    \caption{Settings of IEEE 802.11p MAC used by network nodes in the simulations.}
    \label{tab:ofdmsettings}
    \scriptsize \begin{tabular}{|l|c|}
      \hline
      \textbf{Parameter} & \textbf{Value} \\ \hline
      Slot duration & 0.00013s\\ \hline
      Difs & 0.00032s\\ \hline
      Bitrate & 18 Mbps \\ \hline
      Contention window & 15\\ \hline
      MAC header length & 256 bits \\ \hline
    \end{tabular}
\end{table}

In the experiments using RAdNet-VE and RAdNet, we used active prefixes composed by eight with eight possibilities. Each node created a random prefix once within the network. Interests were defined by the applications implemented to construct the use scenarios. To address the nodes of the CCNs, we used the node identification provided by Omnet++.

In each scenario, the applications exchange large amounts of delay-sensitive data to extract accurate information regarding the number of incoming or outgoing vehicles in upstream segments of the intersections (scenario 1) or information regarding the road conditions and the state of neighboring vehicles (scenario 2). The validity of the data exchanged by these applications is highly time-dependent. The characteristics of these applications show that they must provide non-cacheable data services \cite{Yu:2013}.

\begin{table}[t]
    \centering
    \caption{Interests used by applications of the use scenarios in the simulations experiments using RAdNet-VE}
    \label{tab:interests}
    \scriptsize \begin{tabular}{|>{\centering\arraybackslash}p{8.5cm}|>{\centering\arraybackslash}p{5.0cm}|>{\centering\arraybackslash}p{1.5cm}|>{\centering\arraybackslash}p{1.8cm}|}
    \hline
     \centering \textbf{Data Name} & \textbf{Interests} & \textbf{Maximum Number of Hops} & \textbf{Propagation Directions}\\
    \hline
      \multirow{2}{*}{\textit{uri://vehicle/geolocation/da/vehicle\_entering\_in\_$<$roadId$>$}} 		&  vehicle\_entering\_in\_$<$roadId$>$\_req & 5 & -1\\ \cline{2-4} 
        		   																				&  vehicle\_entering\_in\_$<$roadId$>$\_data & 5 & 1\\ \hline
                   
      \multirow{2}{*}{\textit{uri://vehicle/geolocation/da/vehicle\_leaving\_$<$roadId$>$}} 				&  vehicle\_leaving\_$<$roadId$>$\_req & 5 & 1 \\ \cline{2-4} 
        		   																				&  vehicle\_leaving\_$<$roadId$>$\_data & 5 & -1\\ \hline
                   
      \multirow{2}{*}{\textit{uri://semaphore/geolocation/tlc/ack\_vehicle\_entering\_in\_$<$roadId$>$}} &  ack\_vehicle\_entering\_in\_$<$roadId$>$\_req & 5 & 1\\ \cline{2-4} 
        		   																				&  ack\_vehicle\_entering\_in\_$<$roadId$>$\_data & 5 & -1\\ \hline
      
      \multirow{2}{*}{\textit{uri://semaphore/geolocation/tlc/ack\_vehicle\_leaving\_$<$roadId$>$}} 		&  ack\_vehicle\_leaving\_$<$roadId$>$\_req & 5 & 1\\ \cline{2-4} 
        		   																				&  ack\_vehicle\_leaving\_$<$roadId$>$\_data & 5 & -1\\ \hline
      \multirow{2}{*}{\textit{uri://roadsidesig/geolocation/on/obstacle}} 						&  obstacle\_req & 50 & 1\\ \cline{2-4} 
        		   																				&  obstacle\_data & 50 & -1\\ \hline
                   
      \multirow{2}{*}{\textit{uri://vehicle/geolocation/cacc/presence}} 						&  presence\_req & 1 & 0\\ \cline{2-4} 
        		   																				&  presence\_data & 1 & 0\\ \hline
      \multirow{2}{*}{\textit{uri://vehicle/geolocation/cacc/state}} 							&  state\_req & 1 & 1\\ \cline{2-4} 
        		   																				&  state\_data & 1 & -1\\ \hline                                    
    \end{tabular}
\end{table}

\begin{table}[!t]
    \centering
    \caption{Data names used by applications of the use scenarios in the simulations experiments using CCN$_R$ and CCN$_P$}
    \label{tab:datanames}
    \scriptsize \begin{tabular}{|>{\centering\arraybackslash}p{1.4cm} |>{\centering\arraybackslash}p{2.0cm}| >{\centering\arraybackslash}p{8.5cm}|p{4cm}|}
    \hline
     \centering \textbf{Use Scenario} & \textbf{Application} & \textbf{Data Name} & \textbf{Description}\\
    \hline
      \multirow{17}{*}{1} & \multirow{8}{*}{DA}& \multirow{3}{*}{\textit{uri://vehicle/geolocation/da/vehicle\_entering\_in\_$<$roadId$>$}}& provides data regarding vehicles entering an upstream segment of intersection \\ \cline{3-4}
      &  		   & \multirow{3}{*}  \textit{uri://vehicle/geolocation/da/vehicle\_leaving\_$<$roadId$>$} & provides data regarding the vehicle that isvehicles leaving an upstream segment of intersection \\ \cline{2-4}
                   & \multirow{9}{*}{TLC}& \multirow{4}{*}{\textit{uri://semaphore/geolocation/tlc/ack\_vehicle\_entering\_in\_$<$roadId$>$}} & provides acknowledgment of data regarding vehicles entering an upstream segment of intersection \\ \cline{3-4}
                   &                     & \multirow{3}{*}{\textit{uri://semaphore/geolocation/tlc/ack\_vehicle\_leaving\_$<$roadId$>$}}      & provides acknowledgment of data regarding vehicles leaving an upstream segment of intersection \\ \cline{1-4}
\multirow{7}{*}{2} & \multirow{2}{*}{ON}                  & \multirow{2}{*}{\textit{uri://roadsidesig/geolocation/on/obstacle}}                  & provides data regarding obstacles on a road segment \\ \cline{2-4}
                   & \multirow{5}{*}{CACC} & \multirow{3}{*}{\textit{uri://vehicle/geolocation/cacc/presence}}                       & provides data regarding the presence of other vehicles in the neighborhood  \\ \cline{3-4}
                   &                     & \multirow{2}{*}{\textit{uri://vehicle/geolocation/cacc/state}}                       & provides data regarding the state of following vehicles \\
    \hline
    \end{tabular}
\end{table}

Finally, we configured the maximum number of hops of the protocols used in the simulation experiments. In scenario 1, we configured RAdNet, CCN$_R$ and CCN$_P$ to allow a maximum of five hops. In scenario 2, we configured RAdNet, CCN$_R$ and CCN$_P$ to operate with a maximum of 50 hops. We configured RAdNet-VE in accordance with the interests used by the applications of each scenario and their maximum hop numbers. These settings are given in Table \ref{tab:interests}. According to the Algorithm \ref{alg:radnetveprotocol}, we are also required to configure the maximum number of hops used by RAdNet-VE when the interest matching is false. In this case, the adopted value was 50.

\subsubsection{Data Naming and Interests}

To configure the applications for CCN$_R$ and CCN$_P$, we adopted the following data naming structure: uri://typeofprovider/geolocation/application/dataservicename.

The first component $type of provider$ defines the entity providing the service, i.e., vehicle, semaphore signal, roadside signal. The $geolocation$ component uses the format  \textit{roadId/direction/section} \cite{Wang:2012a}. This component played a significant role in the simulations using CCN$_R$ and CCN$_P$. The applications used $geolocation$ to filter messages according to the position of the nodes. The $application$ component owns the data provided by the node. The $data service name$ gives the name of the particular data service. Based on this data naming structure, we defined the data used by applications as listed in Table \ref{tab:datanames}. 

To configure the applications for RAdNet-VE and RAdNet, we did not adopt the structure described above. We argue that interests are more abstract than the structure of data names based on Universal Resource Identifiers. We do not need to use well-formed strings to identify the data. For example, for data identified as  \textit{uri://vehicle/geolocation/da/vehicle\_entering\_in\_$<$roadId$>$} in CCNs, we simply use the \textit{vehicle\_entering\_in\_$<$roadId$>$} string. The component \textit{$<$roadId$>$} identifies the upstream segment on which the vehicle is traveling. The design of RAdNet-VE and RAdNet ensures that the nodes only use one type of network message. Therefore, we needed to define two interests for each data name given in Table \ref{tab:datanames}. The Table \ref{tab:interests} lists the definitions of each interest. Thus, in the simulations with RAdNet-VE and RAdNet, messages with interests ending with \textit{req} acted as Interest packets, and those with interests ending with  \textit{data} acted as Data packets. Although Table \ref{tab:interests} lists the maximum number of hops and the propagation direction used by RVEP during the simulation experiments, these settings cannot be used by RP (see Algorithm \ref{alg:radnetprotocol}).

\subsubsection{Traffic Settings}

To build the three-by-three grid and road segments and define the vehicle settings, we used SUMO \cite{SUMO:2013}.In each segment of the grid, we set the maximum speed to 60 km/h. The grid had eight entry points, through which flowed 1500 vehicles/hour for a mean of 600 vehicles during the simulations. As the vehicles traveled through the intersections of the grid, each of the 32 semaphore signals communicated with approximately 40 vehicles. The amber phase of each semaphore signal lasted 5 s.

\begin{table}[h]
    \centering
    \caption{Settings for the Intelligent Driver Model (IDM)}
    \label{tab:idm}
    \scriptsize \begin{tabular}{|c|c|c|}
      \hline
      \textbf{Parameter} & \centering\textbf{Value for Use Scenario 1} & \textbf{Value for Use Scenario 2}\\
      \hline
      Desired speed (v$_0$) & 60 km/h & 80 km/h \\ \hline
      Time Headway (T) & 1.2 s & 1.2 s \\ \hline
      Minimum gap (s$_0$)& 2.0 m & 2.0 m\\ \hline
      Acceleration (a) & 1 m/s$^2$ & 1 m/s$^2$ \\ \hline
      Deceleration (b) & 3 m/s$^2$ & 3 m/s$^2$\\ \hline
    \end{tabular}
\end{table}

To build the road segments, we created a 10 km line and positioned the obstacle at one end. We positioned the road-side units for obstacle notification 500 m before the obstacle. Vehicles 1 km away from the obstacle can directly request and receive data regarding the obstacle on the road, since they are within the  communication range of the radio of the road-side unit. We also set the maximum speed between the start of the road segment and the start of the obstacle region to 80 km/h. The obstacle area had a length of 10 m with a maximum speed of 20 km/h. Finally, the road segment received a flow of 1500 vehicle/hour, totaling a mean of 250 vehicles during the simulations.

To configure the behavior of the vehicles, we adopted the car-following Intelligent Driver Model (IDM). Table 3 shows the IDM settings for the two scenarios.

\section{Analysis of Results}

We compared the performance of RAdNet-VE against those of RAdNet, CCN$_R$, and CCN$_P$. The following evaluation measures were used for comparison:

\begin{itemize}
\item \textbf{Message overhead (MO):} the total number of messages received by nodes (including the destination node and forwarding nodes); 
\item \textbf{Latency of the communications among nodes (LCAN):} the time between sending a message from the network layer of the source node to the message being received by the network layer of a neighbor; 
\item \textbf{Data delivery rate (DDR):} the total amount of data received by destination nodes divided by the amount of data sent by source nodes; 
\item \textbf{Number of hops (NoH):} the number of times that the messages are forwarded by nodes; 
\item \textbf{Range of messages (RoM):} the maximum distance that the messages travel as they are forwarded by nodes; 
\item \textbf{Time of message propagation (ToMP): } the time taken for the message to reach a given distance.

\end{itemize}

\begin{figure}[!t]
	\centering
	\includegraphics[scale=0.65]{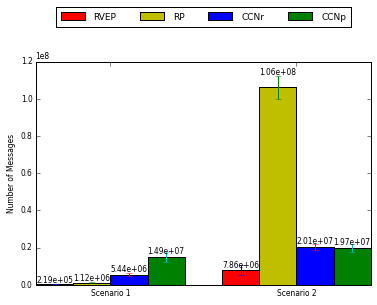}
	\caption{Message overhead (MO)}
	\label{fig:mo}
\end{figure}

\begin{figure}[!t]
	\centering
	\includegraphics[scale=0.65]{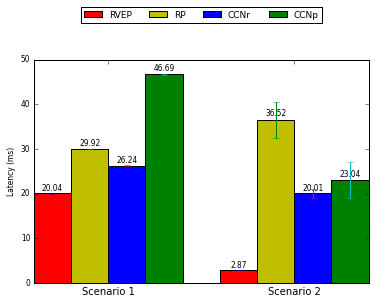}
	\caption{Latency of the communications among nodes (LCAN)}
	\label{fig:lcan}
\end{figure}

\begin{figure}[!t]
	\centering
	\includegraphics[scale=0.65]{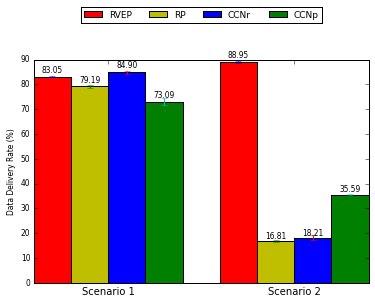}
	\caption{Data delivery rate (DDR)}
	\label{fig:ddr}
\end{figure}

\begin{figure}[!t]
	\centering
	\includegraphics[scale=0.65]{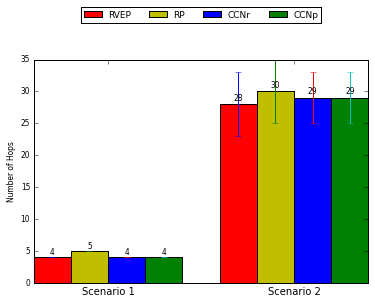}
	\caption{Number of hops (NoH)}
	\label{fig:noh}
\end{figure}

\begin{figure}[!t]
	\centering
	\includegraphics[scale=0.65]{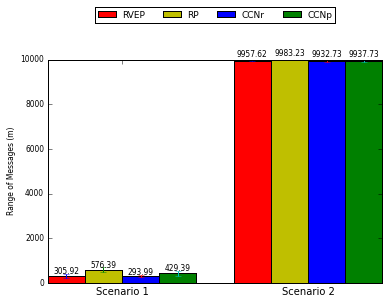}
	\caption{Range of Messages (RoM)}
	\label{fig:rom}
\end{figure}

\begin{figure}[!t]
	\centering
	\includegraphics[scale=0.65]{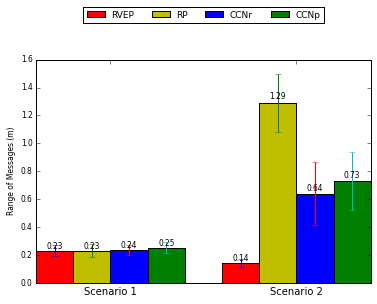}
	\caption{Time of message propagation (ToMP):}
	\label{fig:tomp}
\end{figure}

The results are mean values obtained from the simulation experiments by using RAdNet-VE, RAdNet, CCN$_R$, and CCN$_P$ in the use scenarios. The mean values and standard deviation can be seen in the Table \ref{tab:results}. To properly visualize these results, we present the mean values and standard deviations in the bar charts shown in Figs. \ref{fig:mo}, \ref{fig:lcan}, \ref{fig:ddr}, \ref{fig:noh}, \ref{fig:rom}, and \ref{fig:tomp}. 

Fig. \ref{fig:mo} shows the mean values of the message overheads generated by the protocols in scenarios 1 and 2. Among the networks used in our simulation experiments, RAdNet-VE generated the lowest message overhead. In scenario 1, the mean number of messages generated by RAdNet-VE was 5.12 times lower than that of RAdNet, 24.88 times lower than that of CCN$_R$,and 68.08 times lower than that of CCN$_P$. In scenario 2, the mean number of messages generated by RAdNet-VE was 13.51 times lower than that of RAdNet, 2.55 times lower than that of CCN$_R$, and 2.5 times lower than that of CCN$_P$. The low message overhead of RAdNet-VE is a consequence of its message forwarding mechanism, which uses the values of the position, direction fields and road identifier to filter messages. The low message overhead resulted from its mechanism for registering interests. This mechanism registered interests defined by applications and limited the number of hops which messages containing these interests could reach as nodes forward them. As shown in Fig. \ref{fig:mo}, the benefits of the mechanism for registering interests are more evident in  the scenario 2, since the nodes need only communicate with one-hop neighbors. Thus, the nodes could not broadcast messages beyond the maximum number of hops, which resulted in reduced message overhead, as shown in Fig. \ref{fig:mo}. Due to the low message overhead, messages did not congest the communication channels of the radios in the nodes. 

\begin{table}[!t]
\centering
\caption{Results of the Simulation Experiments}
\label{tab:results}
\scriptsize \begin{tabular}{c|c|c|c|c|c|}
\cline{3-6}
\multicolumn{1}{l}{}                          & 									& \multicolumn{4}{c|}{{\bf Protocols}}                                                                                                                                           \\ \hline
\multicolumn{1}{|c|}{{\bf Use Scenarios}}       & \multicolumn{1}{c|}{{\bf Measures}} & \multicolumn{1}{c|}{{\bf RAdNet-VE}}        & \multicolumn{1}{c|}{{\bf RAdNet}}           & \multicolumn{1}{c|}{{\bf CCN$_R$}}       & \multicolumn{1}{c|}{{\bf CCN$_P$}} \\ \hline
\multicolumn{1}{|c|}{\multirow{6}{*}{{\bf 1}}} & {\bf MO (\textit{msg})}                     & $2.19\times 10^5 \pm 2.83 \times 10^4$   & $1.12 \times 10^6 \pm 1.35 \times 10^5$ &  $5.44 \times 10^6 \pm 9.14 \times 10^5$ & $1.49 \times 10^7 \pm 2.77 \times 10^6$ \\ \cline{2-6} 
\multicolumn{1}{|c|}{}                         & {\bf LCAN (\textit{ms})}                    & $20.04 \pm 0.05$                         & $29.92 \pm 0.05$                        &  $26.24 \pm 0.06$                        & $46.69 \pm 0.08$                        \\ \cline{2-6} 
\multicolumn{1}{|c|}{}                         & {\bf DDR (\%)}                     & $83.05 \pm 0.25$                         & $79.19\pm 0.65$                        &  $84.9 \pm 0.54$                         & $73.09 \pm 1.57$                        \\ \cline{2-6} 
\multicolumn{1}{|c|}{}                         & {\bf NoH (\textit{hops})}                   & 4                                        & 5                                       &  4                                       & 4                                   	  \\ \cline{2-6} 
\multicolumn{1}{|c|}{}                         & {\bf RoM (\textit{m})}                      & 305.92 $\pm$ 79.58                       & 576.39 $\pm$ 113.60                     &  293.99 $\pm$ 42.87		                & 429.39 $\pm$ 110.67                     \\ \cline{2-6}
\multicolumn{1}{|c|}{}                         & {\bf TOMP (\textit{s})}                      & $0.229 \pm 0.040$                       & $0.231 \pm 0.045$                     &  $0.237 \pm 0.037$		                & $0.252 \pm 0.038$                     \\ \hline
\multicolumn{1}{|c|}{\multirow{6}{*}{{\bf 2}}} & {\bf MO (\textit{msg})}                     & $7.86 \times 10^6 \pm 2.53 \times 10^6$  & $1.06 \times 10^8 \pm 6.17 \times 10^6$  & $2.01 \times 10^7 \pm 1.57 \times 10^6$  & $1.97 \times 10^7 \pm 2.09 \times 10^6$ \\ \cline{2-6} 
\multicolumn{1}{|c|}{}                         & {\bf LCAN (\textit{ms})}                    & $2.87 \pm 0.1$                          & $36.52 \pm 4.0$     					 & $20.01 \pm 1.0$                  		& $23.04 \pm 4.0$                       \\ \cline{2-6} 
\multicolumn{1}{|c|}{}                         & {\bf DDR (\%)}                     & $88.95 \pm 0.4$                          & $16.81 \pm 0.4$                        & $18.21 \pm 1.0$                         & $35.59 \pm 0.15$                        \\ \cline{2-6} 
\multicolumn{1}{|c|}{}                         & {\bf NoH (\textit{hops})}                   & $28 \pm 5$                         & $30 \pm 5$                        & $29 \pm 4$                         & $29 \pm 4$                         \\ \cline{2-6} 
\multicolumn{1}{|c|}{}                         & {\bf RoM (\textit{m})}                      & $9957.62 \pm 7.87$                     & $9983.23 \pm 14.34$                    & $9932.73 \pm 28.14$                      & $9937.73 \pm 37.78$                    \\ \cline{2-6} 
\multicolumn{1}{|c|}{}                         & {\bf TOMP (\textit{s})}                      & $0.14 \pm 0.030$                     & $1.29 \pm 0.21$                    & $0.64 \pm 0.226$                      & $0.73 \pm  0.209$                    \\ \hline
\end{tabular}
\end{table}

Fig. \ref{fig:lcan} shows the mean latencies of communication among nodes in scenarios 1 and 2. Among the networks used in our simulation experiments, RAdNet-VE established communications among nodes with the lowest latency values. In scenario 1, the mean latency of communication among RAdNet-VE nodes was 33.02\%, 23.62\%, and 57.07\% lower than that of RAdNet, CCN$_R$ and CCN$_P$, respectively. In scenario 2, the mean latency of communication among RAdNet-VE nodes was 12.72 times, 6.98 times, and 8.01 times lower than that of RAdNet, CCN$_R$ and CCN$_P$, respectively. The low communication latency among RAdNet-VE nodes resulted from low message overhead.

Fig. \ref{fig:ddr} shows the mean values of the data delivery rates in scenarios 1 and 2. In the scenario 1, CCN$_R$ was the most efficient network, although the difference between the mean data delivery rates for CCN$_R$ and RAdNet-VE was only 1.4\%. Regardless of this, we argue that RAdNet-VE is better than CCN$_R$ in scenario 1, as mean message overhead and mean communication latency among nodes were better than those of CCN$_R$. Comparing the mean values of the delivery data rate of RAdNet-VE with those of RAdNet and CCN$_P$, we see that the former yielded gains of 5.06\% and 13.62\%, respectively. In scenario 2, RAdNet-VE was more efficient than the other networks. In this sense, our network generated values 5.29 times, 4.88 times, and 2.49 times better than RAdNet, CCN$_R$, and CCN$_P$, respectively. The excellent data delivery rates of RAdNet-VE are a result of the low message overhead and low communication latency among nodes.

Fig. \ref{fig:noh} shows the maximum number of hops reached by messages as nodes forwarded them during the simulation experiments of scenarios 1 and 2. In scenario 1, only RAdNet achieved five hops, with the protocols reaching four hops. This is possible because RAdNet forwarded all messages that passed through its matching filter (see Algorithm \ref{alg:radnetprotocol}) with a non-zero number of hops. In scenario 2, RAdNet-VE reached the maximum number of hops, better than those of RAdNet, CCN$_R$ and CCN$_P$. Despite this result, the difference between the maximum numbers of hops of the RAdNet-VE and those of the other networks was small, as shown \ref{fig:noh}. 

Fig. \ref{fig:rom} shows the mean range of messages in scenarios 1 and 2. In scenario 1, the messages transmitted by RAdNet-VE nodes reached a mean range of 305.92 m. This shows that the messages remained within the scope of communication of the upstream segments. This is an excellent result, since RAdNet-VE nodes forwarded messages along the upstream segments, and these messages did not cross the intersections. This result is a consequence of the use of road identifiers as one of mechanism for defining a communication scope.  As shown in Fig. \ref{fig:rom}, the messages for CCN$_R$ also lay within the communication scope of the upstream segments. In contrast, the messages for RADNet and CCN$_P$ did not lie within the communication scope of the upstream segments. In scenario 2, the protocols generated very close mean values, as shown in the Fig. \ref{fig:rom}. These results show that RAdNet-VE, RAdNet, CCN$_R$, and CCN$_P$ can attain longer distances.

Fig. \ref{fig:tomp} shows the time taken for messages to propagate a given distance. To obtain the mean values shown in Fig. 6, we measured the required time for a message to travel 300 m in scenario 1 and 10000 m in scenario 2. As shown in Fig. \ref{fig:tomp}, the networks performed very similarly. The differences among the results for the networks were approximately 1 ms. However, in scenario 2, RAdNet-VE yielded a better performance than RAdNet, CCN$_R$ and CCN$_P$, as shown in Fig. \ref{fig:tomp}. Its maximum time for message propagation was 9.12 times lower than that of RAdNet, 4.52 times lower than that of CCN$_R$ and 5.16 times lower than that of CCN$_P$. This is due to the low latency of communication among nodes in RAdNet-VE (see Fig. \ref{fig:lcan}).

\section{Discussion}

In this section, we discuss the communication requirements of applications for VANETs \cite{Willke:2009} and the results obtained from our simulation experiments. 

When an application needs to communicate at low latency, the communication protocol must allow nodes to communicate with low delay \cite{Willke:2009}. According to the results presented in the previous section, RVEP provides communication with low delay. The nodes of RAdNet-VE do not suffer under topological changes in the network, because the mechanism of interest-centric communication does not take network topology into account. Therefore, RVEP does not use control messages to maintain and update data when routing messages between source and destination nodes. Thus, it enjoys a reduced message overhead, and does not need to consume any of the communication bandwidth with control messages.

Regarding the delivery rate, applications need a protocol that can deliver messages to a group of nodes \cite{Willke:2009}. This protocol should ensure a high probability of message delivery \cite{Willke:2009}. The RVEP satisfies part of this requirement through its mechanism of interest-centric communication inherited from the communication protocol of RAdNet \cite{Dutra:2012}. Moreover, the protocol must ensure a high probability of message delivery \cite{Willke:2009}. Therefore, we extended the original message header of RAdNet by adding position, direction and fields. Based on these two fields, the RVEP forward messages by using the nodes farthest from the message source and propagation directions defined by the relevant applications. This results in a reduced message overhead and avoids causing broadcast storm. Another benefits of using position and direction fields is the significantly reduced propagation time over long distances. Thus, the RVEP provides scalability in terms of the distances traveled by messages and the density of the vehicles on road segments. The use of road identifiers was also positive, because these identifiers allow to virtualize the ways, making them a virtual channel for network message transmission.

Applications for controlling individual and group movements (for example, applications of use scenario 2) operate in a well-defined communication scope, which can be a neighborhood of vehicles or a small region within a given network. Therefore, the communication protocol must ensure selective message delivery, which can be based on trajectory, vehicle proximity, or identification of the vehicle  \cite{Willke:2009}. Based on the registration of the maximum number of hops with interests and road identifiers, the RVEP limit the communication scope.

Regarding a low message overhead for membership services, applications similar to those for controlling group movement need a protocol that allows them to maintain and update persistent group structures \cite{Willke:2009}. Since RAdNet-VE is an interest-centric network, the RVEP does not need to address spaces, or centralized and distributed mechanisms, to manage groups. RAdNet-VE is based on the Publisher/Subscriber model. Therefore, through the interest mechanism in the message, it implements membership services in a distributed manner and without high message overhead.

\section{Conclusions}

In this study, we proposed am information-centric network called RAdNet-VE. The results of simulations to test our network showed that RAdNet-VE is a feasible alternative to CCNs, since these networks struggle with a higher message overhead and higher communication latencies among nodes in scenarios where applications need to cooperate in order to achieve some goal. Moreover, the results also showed that RVEP satisfies the communication requirements of applications for VANETs by providing low communication latency, high delivery rates, scalability, a well-defined communication scope, and low message overhead for membership services.

\section*{Acknowledgements}

Acknowledgements to the Conselho Nacional de Desenvolvimento Científico e Tecnológico (CNPq) by financing this research.

\bibliography{main}
\bibliographystyle{plain}

\end{document}